\documentclass[journal,]{./templates/IEEEtran}
\newcommand{\pdftitle}{Radio Access Technology Characterisation Through Object Detection
}

\usepackage[nolist,nohyperlinks]{acronym}
\usepackage{mathtools}
\usepackage{breqn}
\usepackage{amsmath}
\usepackage{balance}
\usepackage{cite}
\usepackage{framed}
\usepackage{siunitx}
\usepackage{soul}
\usepackage{multirow}
\usepackage{etoolbox}
\usepackage{verbatim}
\usepackage{graphicx}
\usepackage{subcaption}
\usepackage{color}
\usepackage[table,xcdraw]{xcolor}
\usepackage{float}
\graphicspath{{./images/}}

\usepackage{url}

\newtoggle{authorcomment}
\toggletrue{authorcomment}

\begin{document}
\bstctlcite{IEEEexample:BSTcontrol}

\title{\pdftitle}
\author{
  \IEEEauthorblockN{
    Erika Fonseca*,
    Joao F. Santos*, Francisco Paisana$^\dagger$,
  and~Luiz~A. DaSilva$^\ddag$
  } 
  
  \IEEEauthorblockA{*CONNECT / the telecommunications research centre \\
    *Trinity College Dublin, Ireland\\
    $\dagger$ Software Radio Systems \\
    $^\ddag$Commonwealth Cyber Initiative, Virginia Tech, USA \\
   \textit{E-mail: \{fonsecae, facocalj\}@tcd.ie,
   francisco.paisan@softwareradiosystems.com,
   ldasilva@vt.edu}
  } \\ 
} 

\maketitle


\begin{acronym}[IMT-Advanced]
  \acro{NR-U}{New Radio Unlicensed}
  \acro{IoU}{intersection over union}
  \acro{mAP}{mean Average Precision}
  \acro{LSTM}{long short term memory}
  \acro{FNN}{fully connected neural network}
  \acro{RForest}{Random Forest}
  \acro{3GPP}{3rd Generation Partnership Project}
  \acro{ABS}{Almost Blank Subframes}
  \acro{Adam}{Adaptive Moment Optimisation}
  \acro{ADC}{Analogue-to-Digital Converter}
  \acro{AMPS}{Advanced Mobile Phone System}
  \acro{AoA}{Angle of Arrival}
  \acro{AoD}{Angle of Departure}
  \acro{AP-CNN}{Amplitude and Phase shift \ac{CNN}}
  \acro{ASP}{Antenna Scan Period}
  \acro{BBU}{Baseband Unit}
  \acro{BER}{Bit Error Rate}
  \acro{BLER}{Block Error Rate}
  \acro{BPSK}{Binary \ac{PSK}}
  \acro{BS}{Base Station}
  \acro{bw}{bandwidth}
  \acro{CBRS}{Citizens Broadband Radio Service}
  \acro{CDMA}{Code Division Multiple Access}
  \acro{CDM}{Code Division Multiplexing}
  \acro{CFO}{Carrier \acl{FO}}
  \acro{C-MTC}{Mission-Critical \acs{MTC}}
  \acro{CN}{Core Network}
  \acro{CNN}{Convolutional Neural Network}
  \acro{CP}{Cyclic Prefix}
  \acro{C-RAN}{Cloud-\ac{RAN}}
  \acro{CR}{Cognitive Radio}
  \acro{CriC}{Critical Communication}
  \acro{CSAT}{Carrier Sense Adaptive Transmission}
  \acro{CS}{Cyclic Suffix}
  \acro{CSMF}{Communication Service Management Function}
  \acro{CV}{Computer Vision}
  \acro{DAC}{Digital-to-Analogue Converter}
  \acro{D-AMPS}{Digital \acs{AMPS}}
  \acro{DC}{Direct Current}
  \acrodefplural{RAT}[RATs]{Radio Access Technologies}
  \acrodefplural{SDS}[SDSs]{Software-defined Switches}
  \acro{DEQUE}{Double-Ended Queue}
  \acro{DL}{Deep Learning}
  \acro{DNN}{Deep Neural Network}
  \acro{DSA}{Dynamic Spectrum Access}
  \acro{DS-CDMA}{Direct Sequence \acs{CDMA}}
  \acro{E2E}{end-to-end}
  \acro{ECC}{Electronic Communications Committee}
  \acro{EDGE}{Enhanced Data rates for \acs{GSM} Evolution}
  \acro{eMBB}{Enhanced \acl{MBB}}
  \acro{eV2X}{Enhanced \ac{V2X}}
  \acro{EV-DO}{Evolution-Data Optimized}
  \acro{FCC}{Federal Communications Commission}
  \acro{FD}{Frame Duration}
  \acro{FDMA}{Frequency Division Multiple Access}
  \acro{FDM}{Frequency Division Multiplexing}
  \acro{FHSS}{Frequency-hopping spread spectrum}
  \acro{FI}{Frame Interval}
  \acro{FM}{Frequency Modulation}
  \acro{FO}{Frequency Offset}
  \acro{FPGA}{Field-Programmable Gate Array}
  \acro{FS}{Flow Space}
  \acro{FSK}{Frequency-Shift Keying}
  \acro{GLDB}{Geo-Location Database}
  \acro{GP-KNN}{Genetic Programming with K-Nearest Neighbors}
  \acro{GPP}{general purpose process}
  \acro{GPRS}{General packet radio service}
  \acro{GPU}{Graphics Processing Unit}
  \acro{GRC}{Global Radio Coordinator}
  \acro{GSM}{Global System for Mobile communication}
  \acro{H2H}{Human-to-Human}
  \acro{HetNet}{Heterogeneous Network}
  \acro{HSDPA}{High Speed Downlink Packet Access}
  \acro{HSPA}{High Speed Packet Access}
  \acro{HSUPA}{High-Speed Uplink Packet Access}
  \acro{iDEN}{Integrated Digital Enhanced Network}
  \acro{IFD}{Inter-Frame Duration}
  \acro{IMT-2000}{International Mobile Telecommunications-2000}
  \acro{IMT-Advanced}{International Mobile Telecommunications-Advanced}
  \acro{INR}{Interference-to-Noise Ratio}
  \acro{IoT}{Internet of Things}
  \acro{IPM}{Intra-Pulse Modulation}
  \acro{ISM}{Industrial, Scientific and Medical}
  \acro{ITU}{International Telecommunication Union}
  \acro{LBT}{Listen Before Talk}
  \acro{LFM}{Linear Frequency Modulation}
  \acro{LRC}{Local Radio Controller}
  \acro{LTE-A}{\acs{LTE}-Advanced}
  \acro{LTE}{Long-Term Evolution}
  \acro{LTE-U}{\ac{LTE} in unlicensed spectrum}
  \acro{LWA}{\acs{LTE}-\acs{WLAN} aggregation}
  \acro{mAP}{mean average precision}
  \acro{MCD}{Measurement Capable Device}
  \acro{mIoT}{Massive \ac{IoT}}
  \acro{ML}{Machine Learning}
  \acro{M-MTC}{Massive \acs{MTC}}
  \acro{MNO}{Mobile Network Operator}
  \acro{MTC}{Machine Type Communication}
  \acro{MVNO}{Mobile Virtual Network Operator}
  \acro{NMT}{Nordic Mobile Telephone}
  \acro{NSMF}{Network Slice Management Function}
  \acro{NS}{Network Slice}
  \acro{NSSMF}{Network Slice Subnet Management Function}
  \acro{OFDMA}{Orthogonal Frequency Division Multiple Access}
  \acro{OFDM}{Orthogonal Frequency Division Multiplexing}
  \acro{ONF}{Open Network Foundation}
  \acro{OS}{Operational System}
  \acro{OTT}{Over-The-Top}
  \acro{PDC}{Personal Digital Cellular}
  \acro{PER}{Packet Error Rate}
  \acro{PLC}{Process Logic Controller}
  \acro{POCSAG}{Post Office Code Standardization Advisory Group}
  \acro{PRI}{Pulse Repetition Interval}
  \acro{PSD}{Power Spectral Density}
  \acro{PSK}{Phase-Shift Keying}
  \acro{PW}{Pulse Width}
  \acro{QoE}{Quality of Experience}
  \acro{QoS}{Quality of Service}
  \acro{RAN}{Radio Access Network}
  \acro{RAT}{Radio Access Technology}
  \acro{ReLU}{Rectified Linear Unit}
  \acro{REM}{Radio Environment Map}
  \acro{RF}{Radio Frequency}
  \acro{RMSE}{Root Mean Squared Error}
  \acro{RMSProp}{Root Mean Square Propogation}
  \acro{RSSI}{Received Signal Strength Indicator}
  \acro{Rx}{receiver}
  \acro{S/A}{Sensors/Actuators}
  \acro{S-CNN}{Spectrogram \ac{CNN}}
  \acro{SDN}{Software Defined Network}
  \acro{SDR}{Software-Defined Radio}
  \acro{SDS}{Software-defined Switch}
  \acro{SER}{Symbol Error Rate}
  \acro{SFI}{Science Foundation Ireland}
  \acro{SIMD}{Single Instruction Multiple Data}
  \acro{SINR}{Signal-to-Interference-plus-Noise Ratio}
  \acro{SNR}{Signal-to-Noise Ratio}
  \acro{SRO}{Symbol Rate Offset}
  \acro{SU}{Secondary User}
  \acro{SVM}{Support Vector Machine}
  \acro{TACS}{Total Access Communications System}
  \acro{TDMA}{Time Division Multiple Access}
  \acro{TDM}{Time Division Multiplexing}
  \acro{TVWS}{TV White Space}
  \acro{Tx}{transmitter}
  \acro{UE}{User Equipment}
  \acro{UHD}{\acs{USRP} Hardware Driver}
  \acro{UMTS}{Universal Mobile Telecommunications System}
  \acro{URLLC}{Ultra-Reliable Low Latency Communication}
  \acro{USRP}{Universal Software Radio Peripheral}
  \acro{V2X}{Vehicular-to-Everything}
  \acro{VM}{Virtual Machine}
  \acro{VOC}{Visual Object Classes}
  \acro{WCDMA}{Wideband Direct Sequence \acs{CDMA}}
  \acro{WiMax}{Worldwide Interoperability for Microwave Access}
  \acro{WLAN}{Wireless Local Area Network}
  \acro{WMN}{Wireless Mesh Network}
  \acro{WMWG}{Wireless and Mobile Working Group}
  \acro{WNV}{Wireless Network Virtualization}
  \acro{WSN}{Wireless Sensor Network}
  \acro{YOLO}{You Only Look Once}
   \acro{5G}{fifth generation of wireless technology}

\end{acronym}

\begin{abstract}
\ac{RAT} classification and monitoring are essential for efficient coexistence of different communication systems in shared spectrum. Shared spectrum, including operation in license-exempt bands, is envisioned in the \ac{5G} standards (e.g., 3GPP Rel. 16).
In this paper, we propose a \ac{ML} approach to characterise the spectrum utilisation and facilitate the dynamic access to it.
Recent advances in \acp{CNN} enable us to perform waveform classification by processing spectrograms as images. In contrast to other \ac{ML} methods that can only provide the class of the monitored \acp{RAT}, the solution we propose can recognise not only different \acp{RAT} in shared spectrum, but also identify critical parameters such as inter-frame duration, frame duration, centre frequency,  and signal bandwidth by using object detection and a feature extraction module to extract features from spectrograms. 
We have implemented and evaluated our solution using a  dataset of commercial transmissions, as well as in a \ac{SDR} testbed environment. The scenario evaluated was the coexistence of WiFi and LTE transmissions in shared spectrum. Our results show that our approach has an accuracy of 96\%  in the classification of \acp{RAT} from a dataset that captures transmissions of regular user communications. It also shows that the extracted features can be precise within a margin of 2\%, 
and is capable of detect above 94\% of objects under a broad range of transmission power levels and interference conditions.  
\end{abstract}

\acresetall


\section{Introduction}
\label{sec:intro}

Spectrum monitoring is necessary for efficient coexistence in shared spectrum; it is also needed for regulators to be able to enforce spectrum policy and identify possible violations~\cite{peha2009sharing}. 
Most of the existing works on spectrum monitoring employ the \ac{RSSI} or energy detection-based methods for detecting
the presence of a signal in the channel of interest \cite{peha2009sharing}
\cite{y2015}. However, these approaches may not be effective when multiple
\acp{RAT}
coexist in the same band \cite{selim2017spectrum}.
In order to operate efficiently, i.e., in an interference-aware manner, wireless
devices operating in shared spectrum must identify other radios and
\acp{RAT} present in the same band before communicating.

For example, 5G introduces,
in 3GPP Rel. 16~\cite{release16}, \ac{NR-U}, with New Radio operating in license-exempt spectrum and therefore required to coexist with other \acp{RAT}.
Even before that, \ac{LTE-U} already must rely on
contextual information about the spectrum usage to operate in shared spectrum~\cite{qualcomm} \cite{outro}.
In this paper, we propose a solution to detect the presence of other \acp{RAT} in the band of interest, classify them, and extract features of the transmission, such as center frequency, bandwidth and duty cycle.

\begin{figure}[t]
  \includegraphics[width=\columnwidth]{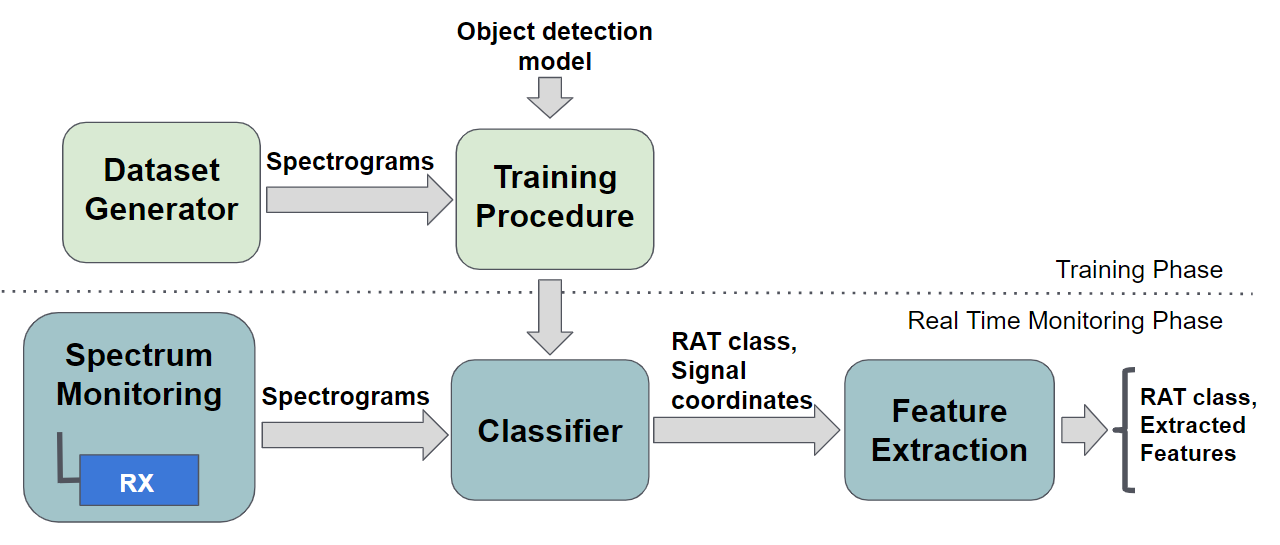}
  \caption{Overview of the proposed approach for characterising \acp{RAT} through object
  detection.}
  \label{fig:pipe}
  \vspace{-1em}
\end{figure}

Our solution employs \ac{DL}, performing object detection
directly on spectrograms, for the characterisation of different \acp{RAT} in
shared spectrum.
We not only classify different \acp{RAT} but also localise the signals in the
frequency and time domains, as well as extract features, including
centre frequency, bandwidth, frame and inter-frame duration. Our model works well under different levels of received signal strength, and in the presence of overlapping transmissions by multiple radios. 

We have also created a dataset generator for training our
\ac{DL} model and evaluating the feature extraction on spectrograms. This was
necessary due to limitations of  publicly available datasets of commercial transmissions,
which do not possess the variations in the frequency and/or time
domains required to evaluate the robustness of the proposed approach.
Our dataset generator automatically produces labelled bounding boxes within
spectrograms, based on the input parameters used in the generation
of the signals. The resulting labelled spectrograms are then used for training
and validating our feature extraction model. Figure \ref{fig:pipe} illustrates
the key components of our proposed approach.


The remainder of this paper is structured as follows. In
Section~\ref{sec:related}, we provide an overview of related work on \ac{ML} applied to
spectrum monitoring. In Section~\ref{sec:classifier}, we introduce our approach
for the characterisation of \acp{RAT}, composed of a \ac{RAT} classifier and
feature extraction components.
In Section \ref{sec:implementation}, we present our dataset generator.
Then, in Section~\ref{sec:validation}, we evaluate the performance of our
classifier and feature extraction under different channel and interference conditions.
Finally, in Section~\ref{sec:conclusion}, we present our concluding remarks and
avenues for future work.

\section{State-of-the-Art on \ac{ML}-based for RF Signal Classification}
\label{sec:related}

Some of the early work on spectrum monitoring has relied on techniques such as cyclostationary
feature detection and energy detection
\cite{paisana2012alternative,ramkumar2009automatic}. 
The focus tended to be on detecting the presence of a signal in the band of interest, rather than characterising the signals being detected.
More recently, there has been renewed interest in modulation classification, driven by spectrum sharing in military bands, and, in the commercial arena, by the possibility of operating 4G and 5G in unlicensed bands, sharing the spectrum with other \acp{RAT} such as WiFi and radar communications.
Current military and commercial spectrum sharing can benefit from more sophisticated awareness of what other transmissions are present in the band, and what the characteristics of those transmissions are, than the earlier spectrum monitoring solutions were able to provide.
This has motivated a number of \ac{ML}-based solutions for signal classification. 

In \cite{wang2017deep}, the authors present several applications
of \acf{DL} algorithms for modulation recognition and channel decoding.
Other works propose the use of \ac{SVM} algorithms \cite{han2017low} or
\ac{GP-KNN} \cite{aslam2012automatic} for modulation classification.
However, both \ac{SVM} and \ac{GP-KNN} techniques are susceptible to frequency
and phase offsets, which can compromise the signal classification accuracy under
multipath, fading, or other real-world \ac{RF} impairments.
Following works have focused on how to make \ac{ML}-based models more robust to \ac{SNR} variations, capturing real-world \ac{RF} impairments and
generating more reliable \ac{RF} signal classifiers. For example, in
\cite{2018over}, the authors generate their dataset through over-the-air
transmissions using \acp{USRP}, evaluating the accuracy of
their classification model under different \acp{SNR}.
The works of \cite{zhang2018dictionary} and \cite{wu2017robust} also consider
different \acp{SNR} when applying \ac{ML} algorithms for modulation
classification. In \cite{zhang2018dictionary}, the authors investigate the
classification problem using dictionary learning, and in \cite{wu2017robust},
the authors utilise \acp{SVM}. In \cite{paisana2017context}, we
characterise the performance achieved by a \ac{CNN} model when identifying distinct
modulations for different \ac{SNR} levels using spectrograms.

The work of \cite{od-oshea} applies computer vision methods to modulation transmission detection. 
They perform object detection on spectrograms using their own generated dataset. 
Their model detects the transmitted frames, but it does not classify them.  
In this paper, we apply object detection to perform not just the detection, but also the classification of different \acp{RAT} and the identification of key features of the transmission. For this, we rely both on a dataset that we generated in the Iris Testbed \cite{doyle2010experiences} and on available datasets of commercial transmissions.

The coexistence between different RATs in shared spectrum requires more information about the surrounding wireless devices than simply the knowledge of their modulation schemes, e.g., QPSK or QAM. For example, different RATs may employ the same type of modulation and yet use different medium access schemes.
In \cite{eletrosense}, the authors propose a \ac{LSTM} model for modulation
classification in large distributed networks of low-cost sensor nodes.
They conclude that a modulation model classifier is not always effective in classifying \acp{RAT}.

The works summarised above provide solutions in the field of signal classification
with, predominantly, high accuracy in what they propose to do. However, most of them are focused on modulation classification and not on \ac{RAT}
classification. These works also do not exploit a scenario with varying degrees of interference and overlapping transmissions among the classified signals, which significantly increases the difficulty in correctly classifying these transmissions. 
In an environment where the spectrum is shared between multiple \acp{RAT} and multiple access points belonging to the same \ac{RAT}, e.g. in the ISM band, transmissions that occur in the same frequency channel can happen with full or partial overlap. Recognising these cases and localising them can lead to better interference and/or coexistence management mechanisms. 

Our earlier work on \ac{RAT} classification~\cite{selim2017spectrum} 
focused on distinguishing between radar, \ac{LTE}, and WiFi transmissions. In \cite{imecpaper} we analyse different machine learning techniques for wireless technology classification with two different datasets, and check the ability of the model to generalise to unforeseen scenarios. These solutions are effective in performing RAT classification but do not provide a more detailed characterisation of the spectrum.

A new approach, presented for the first time in this paper, applies object detection on spectrograms, which allow us to classify and extract key features of the sensed transmissions. The advantage of applying object detection to the RAT characterisation problem is that this technique identifies the object independently of its location in the image, allowing the detection and classification of transmissions that do not occur in the center of the band being observed, with different bandwidths, different duty cycles, etc..
To the best of our knowledge, our work is the first to detect and classify different \acs{RAT} transmissions applying object detection techniques and to evaluate its performance under unfavourable noise and interference conditions. 
Our solution also provides feature extraction functionality, which can be used to build efficient dynamic coexistence mechanisms in shared spectrum. 

\section{\ac{RAT} Characterisation}
\label{sec:classifier}

In this section, we describe our solution for \ac{RAT}
characterisation using object detection. Our approach consists of two main components,  an image-based \ac{RAT} classifier, and post-processing feature extraction, as shown in Figure~\ref{fig:pipe}.
In the following subsections, we describe each of these. 

\subsection{Image-based \ac{RAT} Classifier}

We developed a \ac{CNN}-based classifier for recognising different \acp{RAT}
coexisting in shared spectrum. Our classifier can identify multiple \acp{RAT}
by directly applying object detection to spectrograms. 
The \ac{CNN} must be trained and validated against target
objects. Depending on the size of the neural network and the computing platform available, the training and validation of the \ac{CNN} from scratch may take between
hours and days. 
One way to 
reduce this time is by applying transfer learning,
which relies on the partial reuse of a previously trained model
(trained on a different set of tasks) for addressing a new task. This implies retraining an existing network, typically by fine tuning the weights from the
hidden layers close to the output layer, to make the network more suitable to
the new task. As such, the first layers, which are typically good at extracting
basic features such as edge detection in computer vision tasks, are reused for the new task as well. Transfer
learning significantly decreases the amount of data required for the training
process and, consequently, the duration of the training process.

The application of transfer learning requires the choice of a previously trained network as a starting point. A broad range of pre-trained networks already exists; these are suitable for different problems, e.g., predictive text, speech recognition, and image object detection. 
For the spectrum sharing scenario, where it is necessary to dynamically assess how the spectrum is being occupied, we need a model that can provide acceptable classification accuracy in real-time.
We also require a solution that can provide not just the classification of the object, but also its localisation in the image (as discussed later, we rely on this localisation information for feature extraction). 

  
We employ the well-known object detection model \ac{YOLO} \cite{yolo} as the starting point for our \ac{RAT} classifier. 
\ac{YOLO} is one of the most efficient solutions in the literature for real-time implementation of object detection.
This model outputs both the class of the detected objects, as well as their position in the input image. Using weights and architecture from \ac{YOLO} pre-trained on ImageNet \cite{imagenet}, we modify the Softmax layer, which corresponds to the last layer before the output of the model. During the training process, the Softmax layer is explicitly optimised for the classification of \ac{LTE} and WiFi waveforms. 
The architecture we adopted is presented in \cite{yolo2} and it has 19 convolution layers and 5 max-pooling layers.
Moreover, our model can easily be extended for supporting more \acp{RAT}, by retraining it with datasets that include new waveforms.

\begin{figure}[t]
  \centering
  \begin{subfigure}[b]{0.49\columnwidth}
    \includegraphics[width=\linewidth]{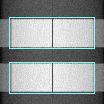}
    \caption{\ac{LTE} detection.}
    \label{fig:1}
  \end{subfigure}
  \hfill 
  \begin{subfigure}[b]{0.49\columnwidth}
    \includegraphics[width=\linewidth]{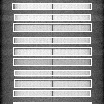}
    \caption{WiFi detection.}
    \label{fig:2}
  \end{subfigure}
  \caption{Spectrogram with the bounding boxes created by our  \ac{ML}-based signal classifier. The positions of the bounding boxes represent the detection of the frame, and the colour represents the classification, blue for LTE and white for WiFi.}\label{fig:spec}
\end{figure}

The training itself requires the fine-tuning of parameters related to the learning rate and convergence of the classification, known as the hyperparameters.
Hyperparameters are parameters chosen before the training process, for example, 
learning rate, optimiser, epochs, etc.   
We detail our choices for the hyperparameters below:

\begin{itemize}

\item {Learning Rate}: is the amount by which the weights in an ML model are updated. We set it to $10^{-5}$; with this value, the model did not overfit and was able to learn the objects' characteristics.

\item {Epoch}: is an iteration of the training process where the model is filled 
with all the elements of the training dataset. 
If a model is trained with too many epochs, it can overfit to the training data, while if a model uses too few epochs, it might not learn the necessary features to perform the classification. 
After testing several values, we set the number of epochs to 50,000.

\item {Mini-batch}: is a part of the dataset used to update the network's weights. The first approaches in \ac{ML} used the entire dataset to update the weights in the network; however, the work of~\cite{masters2018revisiting} argues that this update 
should use a smaller part of the dataset, called a mini-batch. The mini-batch approach can increase model performance when it uses batches with values between 2 to 32 \cite{masters2018revisiting}. In the early stages of the design process of our solution, we observed good performance when setting the mini-batch to 32.

\item{Optimiser}: is the function that modifies the weights of each neuron with the purpose of minimising the loss function. The loss function indicates how close the output of the model is to the expected result. The main objective of the learning process is to optimise the loss function, making the predicted output closer to the expected one without over-fitting to the training data. We chose the optimiser \ac{Adam} because it has the feature of accelerating the search for the minimum value of the loss function and reducing oscillations.

\end{itemize}

After trained, our model produces the identification of the \ac{RAT} (i.e., the result of the classification) and the coordinates of each frame detected in the spectrogram image. 
Figure~\ref{fig:spec} shows examples of LTE and WiFi frames detected, surrounded by bounding boxes: blue for LTE, white for WiFi. 
The four coordinates of each of these bounding boxes are used by the feature extraction component, discussed next.

Once our model is trained and validated, it can provide results on the fly, making it suitable for real time applications.
Our classifier analyses frames in batches of three frames each, providing three outputs at the same time; this allows us to parallelise the classification task and use multiple cores in parallel.
A trade-off that is important to consider is the implication of this design choice on real-time detection and \ac{RAT} classification: the number of images analysed simultaneously cannot be too large,  otherwise the model will not operate in real-time.
 In our implementation, we evaluated the classification speed using an computer with Intel Core i7-6820HK processor and GeForce GTX 1070 Mobile. With this commercial off-the-shelf \ac{GPU}, we are able to analyse three images in around 0.1ms with 2 classes and trained with a commercial transmission dataset (described later).

\subsection{Post-processing Feature Extraction}

Once the classification of the \ac{RAT} is completed, the feature extraction component 
allows us to obtain additional information about the \acp{RAT} present in a given
channel.
The spectrogram corresponds to a band of frequencies [$f_1$, $f_2$], collected during a time interval [$t_1$, $t_2$]. 
Then, we calculate the granularity that each pixel in the image represents in the time and frequency domains, as an increment value in time ($I_T$) and frequency ($I_F$), respectively. This mapping depends on the size of the spectrogram ($[X_{min},X_{max}], [Y_{min},Y_{max}]$)\footnote{Note that uppercase $X$ and $Y$ refer to the spectrogram, and lowercase $x$ and $y$ refer to the bounding box around a frame.}.

\begin{figure}
  \includegraphics[width=\columnwidth]{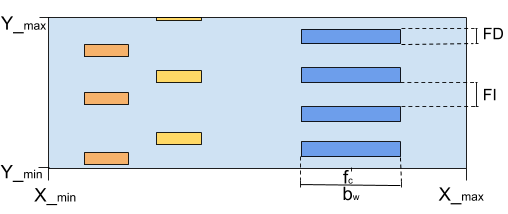}
  \caption{Parameters representation in a spectrogram.}
  \label{fig:post}
  \vspace{-1em}
\end{figure}

The trained model provides the corners of a rectangle that encloses a 
transmission frame, denoted by coordinates $x_{min}, x_{max}, y_{min}, y_{max}$. Given the coordinates of
this rectangle, i.e., the bounding box, as well as the values of each time and frequency increment,
we can localise the signals in the
spectrum and in time. In order to calculate the bandwidth of the signal ($b_w$) and its centre frequency ($f_c$), we use the
horizontal coordinates of the corners of the bounding box, translating them into their respective value in frequency. 
The \ac{FD} of the
signal is calculated in a similar manner, but now using the vertical coordinates of the corners of the bounding box. 
To calculate the average \ac{FI}, we must first calculate the average time
the channel stays without a transmission (CWT), which is the total time
represented in a spectrogram subtracted by the time that is occupied by frame 
transmissions. Then, the \ac{FI} is given by CWT divided by the number of
transmissions on the spectrogram. We summarise the formulas we use for
extracting the features of different \acp{RAT} in Table~\ref{tab:formulas}, and
illustrate the representation of the relevant values on a spectrogram in Figure
\ref{fig:post}.

\begin{table}[]
  \centering
  \caption{The mapping between the image position and the parameters of interest in time and frequency domains.}
  \resizebox{0.49\textwidth}{!}
  {
    \begin{tabular}{|l|l|}
    \hline
    Parameters Time/Frequency & Position Mapping                         \\ \hline
    $I_t$                     & $(t_2-t_1)/(Y_{max}-Y_{min})$                               \\
    $I_f$                     & $(f_2-f_1)/(X_{max}-X_{min})$                 \\
    $b_w$                     & $(x_{max} - x_{min}) * I_f$                \\
    $f_c$                     & $f_1+(I_f*x_{min})+(b_{w}/2)$      \\
    FD                        & $(y_{max} - y_{min})*I_t$                  \\
    CWT                        & $(t_2-t_1) - (frame\_rate * f_{av})$            \\
    FI                        & $CWT/frame\_rate$                         \\ \hline
    \end{tabular}
  }
  \label{tab:formulas}
\end{table}

\section{Dataset Generation} 
\label{sec:implementation}

In the literature, several research efforts
\cite{paisana2017context,kulin2018end} have raised the issue of \ac{RF} dataset
scarcity and had to face the challenge of creating an \ac{RF} dataset before the
development and validation of their \ac{ML}-based classifier.
The lack of labelled \ac{RF} datasets inhibits the development and testing of
new \ac{CNN}-based solutions, and the manual collection of
large \ac{RF} datasets is time-demanding and error-prone.

For the testing and validation of our proposed solution, we have relied on datasets of LTE and WiFi transmissions collected over different locations in Belgium \cite{imecdataset}. However, such datasets of commercial transmissions are not sufficient for the complete evaluation of our feature extraction component.
To evaluate that component it is necessary to have the ground truth for the parameters of the transmissions, as this evaluation is related to the position of the signal in the spectrum and in time. 
In the case of commercial transmissions captured over the air, it is not possible to determine precisely the ground truth, and also it is not possible to vary the \ac{SNR} of the transmission, its centre frequency or its bandwidth, for example. 
To tackle this issue, we designed and implemented a dataset generator for the creation of
labelled \ac{RF} datasets, based on waveforms that mimic the transmissions of LTE and WiFi radios. 
As we are transmitting and receiving the signals, we have full control and knowledge about parameters a priori so that we can generate the ground truth label of the transmissions. This allows the evaluation of the feature extraction that is essential for validating our solution.
We relied on SRS LTE~\cite{srs} for the generation of LTE signals and on a GNURadio implementation \cite{bastian} for the generation of WiFi signals.

\begin{figure}[t]
  \includegraphics[width=\columnwidth]{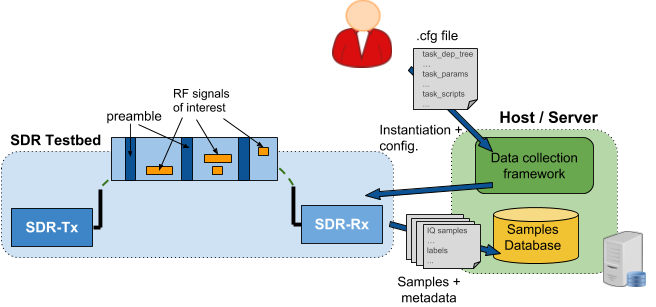}
  \caption{Generating, collecting, and labelling \ac{RAT} transmissions using our dataset generator. The experimenter specifies the waveforms and their
parameters. Then, our \ac{RF} dataset generator creates signal traces with all
the permutations of parameters, as well as transmitting, collecting, and
labelling the signal traces.}
  \label{fig:collection}
\end{figure}

To produce a dataset that reflects real-world transmissions, the dataset must be collected over-the-air to produce samples that undergo RF impairments such as phase/frequency offsets, phase noise, amplifier nonlinearities, etc.  Our dataset generator uses a \ac{SDR} to generate waveforms of different signal strength and bandwidth. It automates the collection and labelling of
over-the-air samples of the waveforms of different \acp{RAT}.
Figure~\ref{fig:collection} depicts the process of generating, collecting and labelling \ac{RF} waveforms using our dataset generator.
These \ac{RF} waveform datasets can be used for training and testing of deep \acp{CNN} for signal classification and spectrum monitoring.

\subsection{Tree of tasks}

Our dataset generator allows the generation of datasets with different: (i)
waveforms, e.g., WiFi, \ac{LTE}, and \ac{PSK} signals; (ii) waveform-specific features, e.g., modulation order and frame length, and DSP transformations, e.g., \ac{FO}, soft gains, shape filtering, and multipath emulation; (iii) \ac{RF} parameters, e.g., centre frequency, hardware gains. Each permutation of parameters and waveform types is translated into IQ signals that are transmitted over the air between \acp{SDR}. Then, the received IQ signals and the associated parameters are stored in data files for later access. 

We developed a
pipeline-based approach for generating traces of \ac{RF} waveforms with
different characteristics. The process is implemented
as a graph of individual tasks, e.g., producing a waveform, setting the frame
duration, and setting the transmission gain. Each task can be configured and run
independently. Each of the task's parameters can be a list of different values,
and the task generates respective output files for all the input values. The
subsequent task receives a set of different input files from the previous task
and performs its operation on all of them. Such a pipeline-based approach
facilitates the extension and inclusion of new tasks, the parallelisation of
tasks, and resuming from intermediate points.

\subsection{Synchronisation and Channel Estimation}
A compelling aspect of our dataset generator is the automatic labelling provided by it, as this is essential for the process of training an \ac{ML} model.
The labelling is created in different formats, including the \ac{VOC} format that is
used in object detection approaches. To provide automatic labelling, it is
essential to keep the \ac{SDR} transmitter and receiver synchronised so that
the labels of their transmitted and received samples remain consistent. We
accomplish the synchronisation and channel estimation through the periodic
transmission of preambles.

The preamble used during dataset generation needs to display strong robustness to noise, so that it can collect samples at the low \ac{SNR} levels that are generally required in \ac{RF} signal/waveform classification use cases. We chose a preamble structure composed of several short Zadoff-Chu sequences with absolute phase shift by an m-sequence for coarse frequency, time offset estimation, and disambiguation, followed by a long Zadoff-Chu sequence for precise frequency offset estimation. For this study, we selected a preamble length of 1031 samples to guarantee robust synchronisation, with probability of preamble detection close to 1 even for values of \ac{SNR} lower than -5 dB. Whenever preamble synchronisation fails, the generator triggers a retransmission.

\section{Performance Evaluation}
\label{sec:validation}
In this section, we evaluate our solution for \ac{RAT} characterisation through
object detection. First, 
we evaluate the detection and classification performance of our model for different RF waveforms under different channel conditions. Next, we assess the feature extraction component of our solution. Then, we estimate the accuracy of the \ac{RAT} classifier and feature extraction components using data from commercial transmissions. 

\subsection{Detection and Classification Performance}
In this section, we evaluate the detection performance and classification accuracy of our model, and demonstrate its robustness in detecting and classifying RF waveforms  under different SNR conditions and interference levels.
We used the dataset generator described in the previous section to compose a dataset of images, i.e., spectrograms and labels, for two radio access technology classes, LTE and WiFi. 
This scenario resembles real world use cases of coexistence in unlicensed spectrum~\cite{wifilte}.
Moreover, our model can be extended, for instance, by increasing the diversity of the RATs included in the training dataset. Extending the training dataset might be useful in a scenario where a technology operating in the unlicensed spectrum might share it with Bluetooth or Zigbee, for example.

\subsubsection{Performance of the Classifier Under Different SNRs}

In this analysis, we evaluate the detection and classification performance of our solution under different SNR conditions. For this evaluation, we generated a dataset with different levels of transmission power, measuring the SNR at the receiver side. We used 400 images to train the model and adopted the configuration described in Section \ref{sec:classifier}, which empirically produced satisfying results. As explained in Section \ref{sec:implementation}, our dataset generator has a minimum SNR threshold value for synchronisation of the preamble over-the-air. The measurements start with an SNR value of -13dB and go up to 35dB. Each spectrogram represents a 50ms time interval and a 20MHz band.

\begin{figure}
  \includegraphics[width=\linewidth]{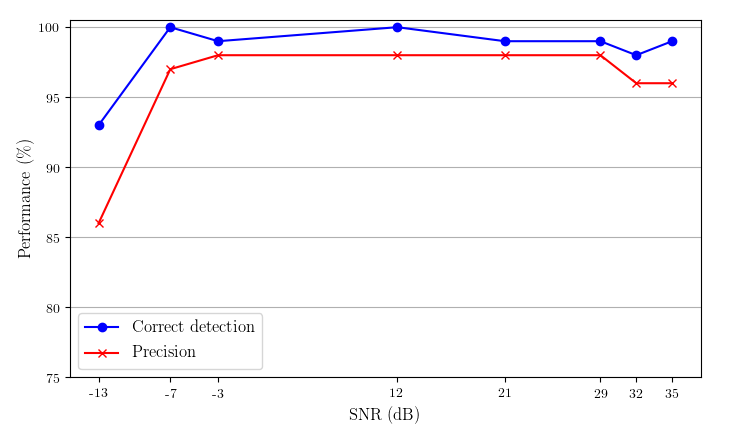}
    \caption{Percentage of correctly detected objects and precision as a function of SNR.}
    \label{fig:snr}
\end{figure}

\begin{figure}[t]
  \centering
  \begin{subfigure}[b]{0.3\columnwidth}
    \includegraphics[width=\linewidth]{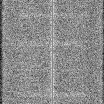}
    \caption{WiFi detection for SNR of -13dB.}
    \label{fig:wifi-02}
  \end{subfigure}
  \hfill 
  \begin{subfigure}[b]{0.3\columnwidth}
    \includegraphics[width=\linewidth]{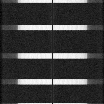}
    \caption{WiFi detection for SNR of 12dB.}
    \label{fig:wifi-05}
  \end{subfigure}
  \hfill 
  \begin{subfigure}[b]{0.3\columnwidth}
    \includegraphics[width=\linewidth]{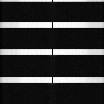}
    \caption{WiFi detection for SNR of 35dB.}
    \label{fig:wifi-099}
  \end{subfigure}
  \caption{Illustration of WiFi signals under different SNRs.}\label{fig:snrs}
\end{figure}

First, we are interested in assessing the ability of our model to detect the transmitted frames correctly.  
The top curve in Figure \ref{fig:snr} shows the percentage of correctly detected frames as a function of SNR.
Detection is around 98\% for all SNR values tested, except -13 dB:
at that SNR, the edges of the transmitted frames are not as sharp, 
as illustrated in Figure~\ref{fig:snrs}, resulting in a lower probability of detection.

Next, we are interested in assessing our model's ability to classify the detected frames. The precision metric is commonly used in classification problems \cite{metrics}, and it represents the percentage of all detected frames that are correctly classified.
The precision is shown in Figure~\ref{fig:snr} and varies from 86\% for an SNR of -~13dB 
to 98\% at SNR between -3 and 32dB. For the highest SNRs, 32 and 35dBs, we obtained an accuracy of 96\%. 
It is worth mentioning that when the SNR is very high the leakage in the transmission also increases, as illustrated in Figure~\ref{fig:snrs}, which in our evaluation compromised 2\% of classification accuracy.

Figure~\ref{fig:snr} shows that when the SNR is low, both the ability to detect the frame and to correctly classify it are impaired. Although the higher leakage does not influence the ability to detect the frames, it slightly affects the classification performance.

\subsubsection{Interfering Transmissions Under Different SNRs}

In this analysis, we evaluate the ability of our model to detect and classify frames under the effect of cross-technology interference. 
We consider two signals with the same bandwidth: the desired signal is an LTE transmission, and the interfering signal is a WiFi transmission.
The desired signal is transmitted with an SNR of 29 dB, and the \ac{SNR} of the interfering signal varies between 3dB to 35dB, both in the same centre frequency and with 20MHz of bandwidth. The spectrograms have the same characteristics mentioned in the previous section.

\begin{figure}
  \includegraphics[width=\linewidth]{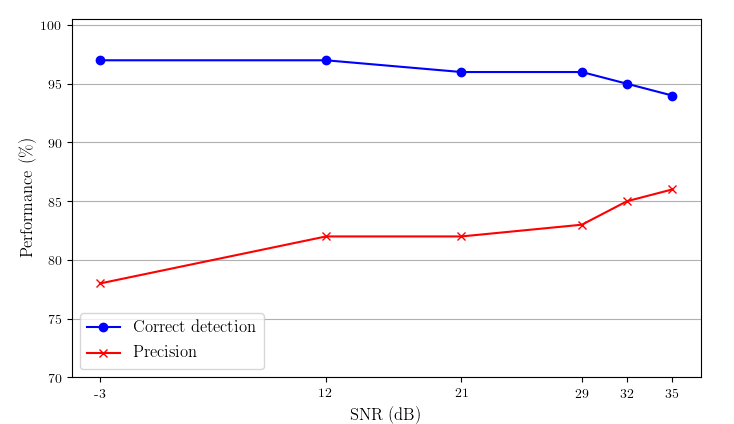}
    \caption{Correct object detection and precision per SNR of the interference signal.}
    \label{fig:over}
\end{figure}

Figure~\ref{fig:over} shows the results of our experiment.
The model could detect the \ac{LTE} frames 97\% of the time, with this accuracy declining slightly as the \ac{SNR} of the interfering WiFi transmission increases.
The curve representing the precision of the model shows that it improves in classifying the frames once the SNR of the WiFi signal increases. This happens because when the interfering WiFi frames had lower \ac{SNR}, the model had issues clearly classifying the transmissions as either \ac{LTE} or WiFi. However, once the SNR of the WiFi is higher than the \ac{SNR} of the LTE transmissions, the model is more successful on classifying them, achieving 86\% of accuracy.

These results show that even in a scenario of strong cross-technology interference, our model is capable of detecting the frames and classifying different RATs, providing a reasonable characterisation of the environment.
To the best of our knowledge, this is the first work to assess the performance of an ML model for RAT classification under the effect of interference with overlapping transmissions. 

\subsection{Feature Extraction}\label{feature}

\begin{figure*}
\begin{subfigure}[h]{0.5\linewidth}
\includegraphics[width=\columnwidth]{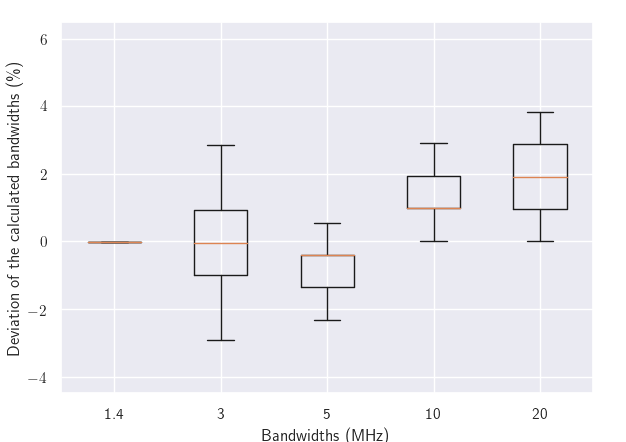}
    \caption{Bandwidth deviation.}
    \label{fig:acband}
\end{subfigure}
\hfill
\begin{subfigure}[h]{0.5\linewidth}
 \includegraphics[width=\columnwidth]{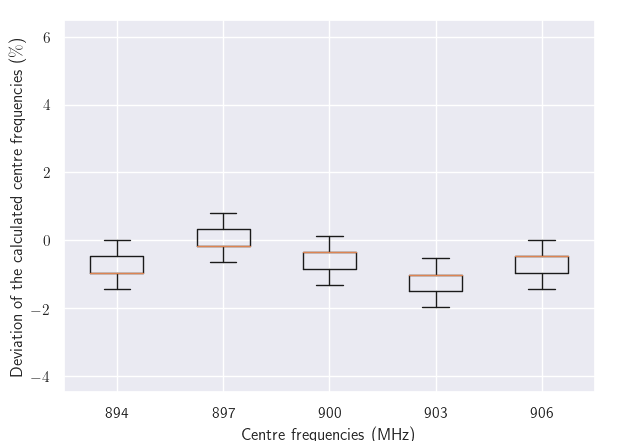}
    \caption{Centre frequency deviation.}
    \label{fig:acfreq}
\end{subfigure}%
\hfill
\begin{subfigure}[h]{0.5\linewidth}
 \includegraphics[width=\columnwidth]{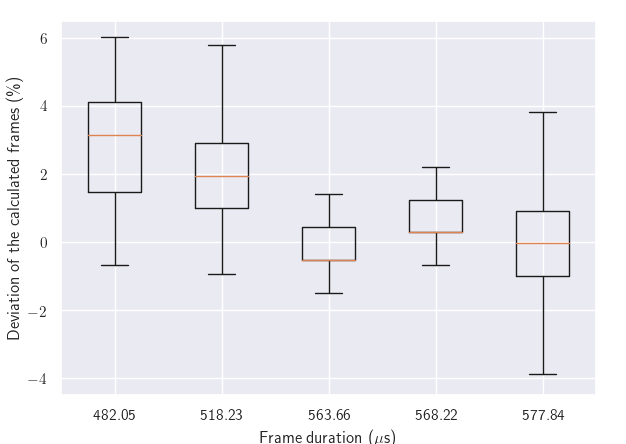}
    \caption{Frame duration deviation.}
    \label{fig:acframe}
\end{subfigure}%
\hfill
\begin{subfigure}[h]{0.5\linewidth}
 \includegraphics[width=\columnwidth]{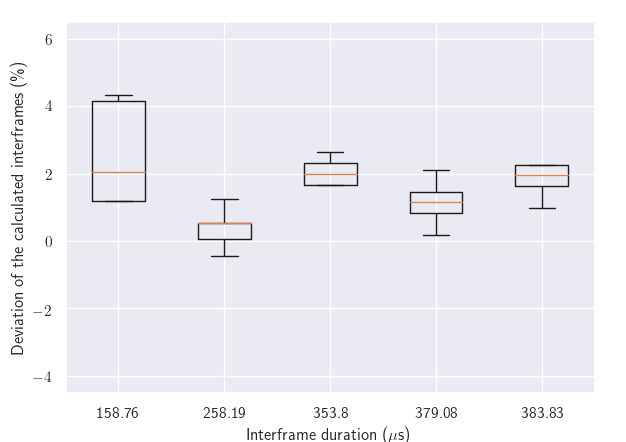}
    \caption{Inter-frame duration deviation.}
    \label{fig:acinter}
\end{subfigure}%
\hfill
\caption{Feature extraction deviation evaluation in time and frequency domain.}
\label{fig:feature}
\end{figure*}

To evaluate the capabilities of our feature extraction component, we generated several datasets using different combinations of: transmission bandwidths, frame duration, inter-frame duration, and centre frequency. 
The average SNR of the transmissions in this evaluation is 29 dB.
Figure~\ref{fig:feature} illustrates the 
accuracy in the feature extraction, for different transmission characteristics.
In our experiments, the value of the $I_f$ is 192.307KHz, which means that each pixel in the spectrograms accounts for a variation of 192.307KHz in the frequency domain. For example, if the calculated centre frequency is off by a single pixel, the computed value will deviate 192.307KHz from the correct centre frequency. The same applies in the time domain, where each pixel accounts for a variation of $I_t=519\mu$s.

Figures~\ref{fig:acband} and~\ref{fig:acfreq} illustrate the accuracy in the extraction of frequency domain features. For all cases tested, the median deviation from the ground truth is  at most 2\%.

The results of the extraction of time-domain features are shown in Figures~\ref{fig:acframe} and~\ref{fig:acinter}. For these, the median deviation from the ground truth is at most 4\%.
Figure~\ref{fig:acframe} illustrates that when the frame duration to be detected is smaller, the solution tends to have an average error higher than when the frame has a longer duration. This happens because it is harder to identify the precise size of smaller objects. 
As depicted in Figure \ref{fig:acinter}, the extraction of inter-frame duration shows similar accuracy.
Our model detects a high  percentage of the transmitted frames. Whenever the model fails to detect a frame, it assumes that the spectrum is empty for that period, increasing the extracted inter-frame duration. However, even in those cases, our model achieves a median deviation of less than 2 percent in all the cases. 

Considering the results discussed in this subsection, we can conclude that our model is capable of extracting the signal features with high precision. Moreover, if necessary for specific applications, a higher precision can be achieved by using higher-resolution spectrograms, i.e., smaller $I_f$ and $I_t$ values.

\subsection{Performance Comparison Using Public Datasets}
In this section, we evaluate our model using a publicly available dataset of commercial LTE and WiFi transmissions collected in Belgium. This evaluation is crucial because it shows that our model can work in real-world scenarios.

First, we investigate the accuracy of our model as a function of the number of spectrograms in the training dataset.
Then, to demonstrate the ability of our object detection model to classify commercial transmissions accurately, we compare our solution to the ones proposed in \cite{imecpaper}, which used the same publicly available dataset. 


\begin{figure}
  \includegraphics[width=\columnwidth]{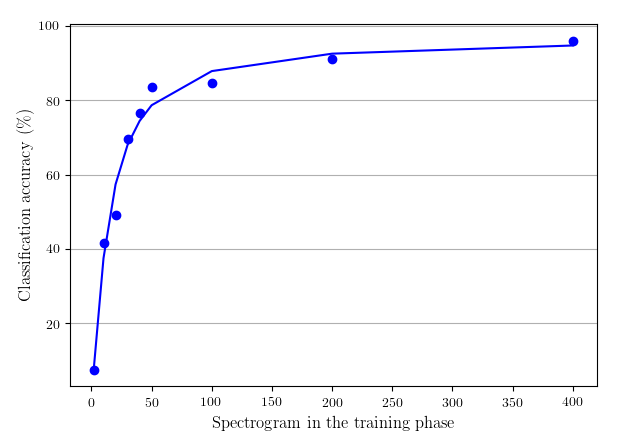}
  \caption{Number of spectrogram in the training phase versus accuracy of the model.}
  \label{fig:specresult}
\end{figure}

We start by analysing how the number of the samples (spectrogram images) affects the performance of the proposed model.
The volume of training data can limit the application of \ac{ML}, because  \ac{ML} techniques usually require a considerable amount of data to learn. 
For example, the work of \cite{imecpaper} used more than 12 thousand images for training the CNN solution based on spectrograms. 
In this section, we assess the performance of our model, considering the volume of training data.

We repeated the training in an identical setup while only adjusting the number of spectrograms used: 2, 10, 20, 30, 40, 50, 100, 200, and 400. The training samples equally represent the LTE and WiFi classes.
Figure~\ref{fig:specresult} illustrates how accuracy depends on the number of spectrograms used in training the model. 
The best accuracy achieved was 96\% with 400 spectrograms.
Hence, we limited the size of our training dataset to 400 images, as this volume of training data is sufficient for our model to achieve a comparable accuracy to the CNN image-based solution presented in \cite{imecpaper}, while using a considerably lower number of training images (only 3.23\% of the dataset size used in \cite{imecpaper}). 


\begin{figure}
  \includegraphics[width=\columnwidth]{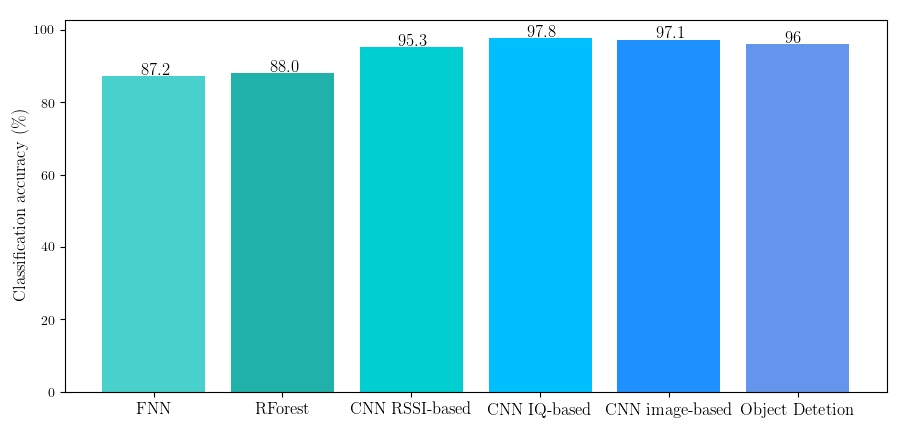}
  \caption{Classification accuracy of different ML solutions.}
  \label{fig:comparison}
\end{figure}

We then compared the object detection-based classification solution presented in this paper against other \ac{RAT} classification solutions in \cite{imecpaper}. These solutions include a \ac{FNN}, a \ac{RForest} \cite{random}, a \ac{CNN} solution based on \ac{RSSI}, a \ac{CNN} solution based on IQ samples, and a CNN solution based on spectrograms. The results of this comparison are shown in Figure \ref{fig:comparison}. 
The CNN-based solutions, including the solution presented in this paper, correctly identify the RAT with  accuracy above 95\%.
The CNNs for IQ and image-based solutions  achieve marginally better accuracy compared to our proposed solution. However, our solution provides additional information regarding spectrum usage that can enhance the efficient use of the spectrum.

\section{Conclusion}
\label{sec:conclusion}
In this paper, we presented an \ac{ML}-based classifier for RAT characterisation using object detection. To the best of our knowledge, this is the first work that evaluates the use of an object detection model for radio technology classification, under multiple interference conditions and employing real user data. Our proposed approach combines the application of object detection on spectrograms for classifying different \acp{RAT} and a feature extraction component for further characterising the \acp{RAT}. From spectrogram images, we can extract specific features from the \ac{RAT}, e.g., inter-frame duration, frame duration, centre frequency, and signal bandwidth. To evaluate the classification accuracy of our model, we trained and classified spectrograms created using the public dataset \cite{imecdataset} that was collected in different locations in Belgium. To evaluate the feature extraction component of our approach, we developed a prototype implementation of the \ac{RAT} classifier in software defined radio. We trained our classifier with \ac{LTE} and WiFi waveforms and showed its efficiency in detecting and classifying different \acp{RAT}. 
Furthermore, we evaluated the resilience of our feature extraction component through transmissions over-the-air with different bandwidths and centre frequencies under distinct SNRs and with overlapping transmissions in the same band.

Our solution
can be useful in spectrum monitoring applications and to facilitate the coexistence of different \acp{RAT} in shared spectrum, as envisioned in 5G. However, we believe that there are improvements to be made in the generation of the labelled data from commercial transmissions. For instance, the process of manually labelling data is time consuming and error-prone. 

The source code of the prototype implementation of our dataset generator can be found on GitHub \cite{gitgenerator} for local use with \acp{USRP}, or remote use in the Iris Testbed \cite{doyle2010experiences}. Our implementation is based on widely used frameworks, such as GNU Radio, for digital signal processing, and YOLO, for real-time object detection. This facilitates the use of our proposed approach by the community and enables further potential applications related to spectrum sensing.

\section*{Acknowledgements}

The research leading to this work received funding from the European
Horizon 2020 Program under the grant agreement No. 732174 (ORCA project). In addition, this work was partly funded by Science Foundation Ireland (SFI) and the National Natural Science Foundation of China (NSFC) under the SFI-NSFC Partnership Programme Grant Number 17/NSFC/5224.


\balance

\bibliographystyle{./templates/IEEEtran}
\bibliography{IEEEabrv,main}

\end{document}